\documentclass[final,5p,times]{elsarticle}

\usepackage{pdflscape}
\usepackage{amssymb}
\usepackage{amsmath}

\usepackage{longtable}
\usepackage{tabularray}
\usepackage{graphicx}
\usepackage{dcolumn}
\usepackage{bm}
\usepackage[utf8]{inputenc}
\usepackage[T1]{fontenc}
\usepackage{mathptmx}
\usepackage{etoolbox}
\usepackage{float}
\usepackage{hyperref}

\usepackage[table,xcdraw]{xcolor}
\usepackage{booktabs, tabularx}
\usepackage{xparse}  
\usepackage{siunitx}
\usepackage{natbib}
\usepackage{amsthm} 
\usepackage[protrusion=true,expansion=true]{microtype} 
\usepackage{placeins}
\usepackage{cuted}
\usepackage[english]{babel}
\usepackage{makecell, multirow, threeparttable}
\usepackage{xcolor}

\usepackage{xparse}    
\usepackage{xstring}
\usepackage{dcolumn}
\usepackage{relsize}
\usepackage{varioref}
\usepackage{hyperref}
\hypersetup{colorlinks = true, allcolors = blue}
\usepackage[nameinlink,noabbrev]{cleveref} 

\journal{Atomic Data and Nuclear Data Tables}

\begin{document}

\begin{frontmatter}



\title{Experimental cross sections for K-shell ionization by electron impact}

\author[1]{S. P. Limandri}

\author[1]{A. C. Carreras}

\author[1]{J. C. Trincavelli}

\author[1]{J. A. Guzmán}

\author[2]{D. M. Mitnik}

\author[2]{C. C. Montanari}

\author[1]{S. Segui\corref{cor1}}
\ead{silvina.segui@mi.unc.edu.ar}

\cortext[cor1]{Corresponding author}

\affiliation[1]{organization={Instituto de Física Enrique Gaviola (IFEG-CONICET), Facultad de Matemática, Astronomía, Física y Computación (FAMAF), Universidad Nacional de Córdoba},
city={Córdoba},
country={Argentina}}

\affiliation[2]{organization={Instituto de Astronomía y Física del Espacio (IAFE-CONICET), Universidad de Buenos Aires},
city={Buenos Aires},
country={Argentina}}

\begin{abstract}

A comprehensive compilation of experimental K-shell ionization cross sections induced by electron impact has been assembled, including results up to December 2024. The data are organized according to the target atomic number and to the incident electron energy for elements ranging from H to U. From the 2509 reported data, more than 50\% pertain only to 8 elements (H, He, Ar, Cr, Fe, Ni, Cu, and Ag). Conversely, 13 elements have only one or two results, and no data is available for 27 elements in the range of atomic numbers considered. Additionally, a further inspection of the database reveals that the majority of the data is concentrated within a small energy range, spanning up to four times the K-shell ionization energy. Finally, the different methods used to measure the ionization cross section are analyzed and a discussion about the main sources of uncertainties is presented.
\end{abstract}




\end{frontmatter}


\date{June 19, 2025}

\section{\label{sec:intro}Introduction}

The ionization cross section $\sigma_K$ for the K shell is a relevant parameter in different fields of physics, such as atomic, plasma, and radiation physics. In particular, for electron incidence, several materials characterization techniques depend on the knowledge of this magnitude; this is the case for electron probe microanalysis (EPMA), Auger electron spectroscopy (AES), and electron energy-loss spectroscopy (EELS).

The experimental determination of inner-shell ionization cross sections by electron impact can be accomplished using different methods: by counting the ions or secondary electrons produced by the incident beam (used exclusively for H and He), or by spectroscopic techniques involving X-ray and Auger emissions, and energy loss experiments~\cite{powell76}. The targets employed can be gaseous or solid; the latter include thin films, either self-supported or deposited on substrates, and bulk samples. Depending on the method used, in some cases the ionization cross section must be derived from the X-ray production cross section ($\sigma_K^X$), which is the product of $\sigma_K$ and the K-shell fluorescence yield $\omega_K$. In general, since the K shell is the innermost atomic level, any dependence of $\sigma_K$ on the molecular environment or the physical state of the target is assumed to be negligible.
 
Despite the great effort devoted to the subject for almost a century, the dependence of $\sigma_K$ on the atomic number $Z$ and the kinetic energy $E$ of the incident electron cannot be fully described. Significant discrepancies are found between sets of experimental determinations from different authors, and between experiment and predictions~\cite{llovet14}. The evaluation of theoretical models requires a comprehensive database, including all the experimental cross-section values reported up to date. In 1990, Long {\it et al.}~\cite{long90} compiled experimental data existing up to December 1989. In a new review, Liu {\it et al.}~\cite{liu00} added data up to December 1999. These reviews tabulate numerical data reported in the original publications and a few data obtained from private communications; they also record the existence of graphical data (not included in the tabulations). Fourteen years later, in an extensive work, Llovet \textit{et al.}~\cite{llovet14} updated the K-shell ionization database with measurements reported up to December 2013, presenting the data in a comprehensive set of plots. It is worth mentioning that these authors also compile data for L- and M-shell ionization and L$\alpha$ X-ray production cross sections, and make comparisons with different theoretical approaches. However, the numerical data are not included in that publication. It is desirable to have an updated database with the cross-section numerical values, including those reported in the last 10 years.

In the present work, an exhaustive search for experimental values of K-shell ionization cross sections by electron impact was performed. A new database was created that includes numerical values, increasing the number of data reviewed in previous compilations from 1953 to 2509. Some particular cases were analyzed, and the experimental values were compared with the theoretical predictions performed by Bote \textit{et al.}\cite{bote09}.
 
\section{\label{experimental}Experimental methods}

There are different experimental methods for determining the K-shell ionization cross section. Some of them are summarized below. 

\subsection{Ions and secondary electrons}

For hydrogen and helium, these magnitudes are obtained from the measurement of the number of H$^+$ or He$^+$ ions, or secondary electrons produced by interactions with the incident beam, since the ionization can only occur in the K shell. In this case, the K-shell ionization cross section can be written as \cite{rapp65}: 

$$\sigma_K=\frac{N}{k\delta'}$$

\noindent where $N$ is the number of ions (or secondary electrons) detected per incident electron, $\delta'$ is the number of target atoms per unit area, and $k$ accounts for the detection efficiency of particles or secondary electrons. This expression is valid for the beam-static-target technique. Instead, for the crossed-beams technique, the previous expression becomes more complicated and includes the beams velocities, and the detection of neutral atoms \cite{montague84}.  The main source of uncertainties in this last case corresponds to the detection efficiency, estimated as $5\%$, which leads to a final uncertainty of around 6-7\% for incident energies between 30 and 750~eV. For the remaining elements, the K-shell ionization cross sections have been determined from measurement of X-ray yields, Auger electron yields, or EELS spectra, using different kinds of targets.

\subsection{X-rays}
In the case of X-rays, the number of detected K-line photons $N_{\rm x}$ can be written as:

\begin{equation}
N_{\rm x} = f(\sigma_K) \, N_{\rm e} \, \varepsilon \, \frac{\Delta \Omega}{4\pi} \, \omega_K \, ,
\label{nx}
\end{equation}

\noindent where the first factor is a function of the K-shell ionization cross section that depends on the type of sample. 
This expression includes the contribution of K$\alpha$ and K$\beta$ lines; if only one of them is considered, it must be multiplied by the corresponding relative transition probability. 

The remaining factors in Eq. \eqref{nx} correspond to the number of incident electrons $N_{\rm e}$, the intrinsic detector efficiency $\varepsilon$, the solid angle fraction subtended by the detector $\Delta \Omega / 4\pi$, and the fluorescence yield $\omega_K$. The uncertainties associated with these parameters have been studied in several situations, and it has been found that the uncertainty in the number of incident electrons is lower than $1\%$, and for the solid angle fraction it is around $5\%$~\cite{limandri12}. Regarding intrinsic detector efficiency $\varepsilon$, it is a function of the photon energy that drops sharply at low energies, and it depends on the nature of the detector. There are several methods to determine it, for example, by using radioactive sources or by comparing measured and predicted spectra. In the case of K lines of light elements ($Z<12$), ultra-thin-window X-ray detectors are typically employed, which exhibit an efficiency with significant jumps at the absorption edges of the elements contained in the window. The resulting uncertainties in the detector efficiency for energy dispersive spectrometers were estimated to be around $10\%$ in the energy range between the C-K and O-K lines. These uncertainties decrease with increasing photon energies, being around 1\% for Al-K$\alpha$ and Si-K$\alpha$ and 0.2\% for Ti-K$\alpha$~\cite{limandri12}.

\begin{figure}[h!]
    \centering
    \includegraphics[width=1.0\linewidth]{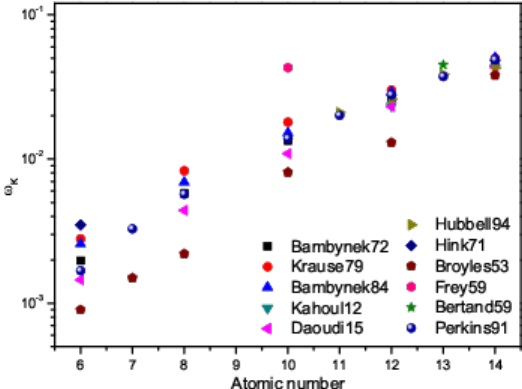}
    \caption{K-shell fluorescence yield for low atomic numbers taken from Bambynek \textit{et al}. \cite{bambynek72,bambynek84}, Krause \textit{et al}. \cite{krause79}, Kahoul \textit{et al}. \cite{kahoul12}, Daoudi \textit{et al}. \cite{daoudi15}, Hubbell \textit{et al}. \cite{hubbell94}, Hink and Paschke \cite{hink71}, Broyles \textit{et al}. \cite{broyles53}, Frey \textit{et al}. \cite{frey59}, Bertrand \textit{et al}. \cite{bertrand59}, and Perkins \textit{et al}. \cite{perkins91}. }
    \label{omegas}
\end{figure}

Regarding the fluorescence yield, the uncertainties in the experimental values for $\omega_K$ range from about
$10\%$ for $Z=11$ to approximately
$1\%$ for $Z=80$ and above \cite{hubbell94}. According to a more recent work \cite{martins20}, the discrepancy between different theoretical, semiempirical and experimental sources is even greater for lower atomic numbers, where $\omega_K$ has a very small value (see Fig. \ref{omegas}).

The function $f(\sigma_K)$ in Eq. \eqref{nx} depends on the kind of target used for the determination of $\sigma_K$. In the case of self-supported thin targets, it is assumed that the incoming electrons follow a linear path within the sample without energy loss. Thus, the following relationship can be written:
\begin{equation}
f(\sigma_K) = \sigma_K \delta t\;,
\label{selfsupported}
\end{equation}
where $\delta$ is the target atomic density (number of atoms per volume unit) and $t$ is its thickness. If the film is not thin enough, corrections to account for non-linear paths and the energy loss suffered by the electrons within the material must be performed.
The main source of uncertainties in Eq. \eqref{selfsupported} is related to the number of atoms per unit area of the film, given by the product $\delta t$. These uncertainties can be estimated between 5\% and 30\%, depending on the characteristics of the film and the technique used to measure its thickness \cite{gil12,bergese06, pazzaglia19}. Films can have problems of uniformity, for example, by clustering of matter. To reduce these effects, films can be supported on light substrates (\textit{e.g.} carbon, mylar or aluminum). In this case, the X-ray enhancement due to the contribution of the electrons backscattered from the substrate must be taken into account.

When diluted gases are used as targets, Eq.\eqref{selfsupported} can also be applied, where the number of atoms per unit area, $\delta t$ 
depends on pressure and temperature, and can be determined with a high degree of precision (2\%) \cite{hoffmann79}. For this kind of target, another method, based on the measurement of characteristic and bremsstrahlung emissions, can be used to cancel out certain experimental parameters \cite{hippler81}, which leads to:
\begin{equation}
N_{\rm x} = \sigma_K \frac{N_{\rm b}}{4\pi} \left(\frac{d^2\sigma_{\rm b}}{d\Omega\, dE}\right)^{-1}\frac{1}{\Delta E} \, \omega_K \, ,
\label{bremss}
\end{equation}
where $N_{\rm b}$ is the measured bremsstrahlung intensity, $\frac{d^2\sigma_{\rm b}}{d\Omega\, dE}$ is the double differential bremsstrahlung cross section, and $\Delta E$ is an energy interval centered at the photon energy. The main source of uncertainties of this method is around 10\%, and corresponds to the bremsstrahlung cross section \cite{kissel83}.

For thick targets, the situation is more complex due to the deflection and energy loss of the electrons within the sample. This problem can be overcome by means of the integral-differential method \cite{an08}, where two assumptions are made: on the one hand, the incident electron describes a linear path within the sample, and on the other hand, the emissions generated by secondary particles are negligible. In this formalism, the function $f(\sigma_K)$ in Eq. \eqref{nx} can be written as
\begin{eqnarray}
\label{NdeR}
 f(\sigma_K) &=& \frac {\omega_KN_0}{A} 
  \int_0^R \sigma_{K}(E(\rho x)) \nonumber \\
  &&\exp \left[ -\mu_{\rm x} \frac{\cos\alpha}{\cos\beta}
   \int^{\rho x}_0  \,{d(\rho x')} \right]  \,{d(\rho x)}
\end{eqnarray}
where $N_0$ is the Avogadro number, $A$ is the atomic mass, $R$ is the range for the electrons with energy enough to ionize the level of interest, $\rho x$ is the mass length traveled by the electrons, $\mu_{\rm x}$ is the mass attenuation coefficient for the characteristic X-rays inside the target, and $\alpha$ and $\beta$ are, respectively, the incident and exit angles measured with respect to the normal of the sample surface. In terms of the stopping power $S(E)=- dE/d(\rho x)$, and using the Leibniz’s rule for differentiation, 
$\sigma_{K}$ can be cleared from equation Eq. \eqref{NdeR} \cite{perez15}:

\begin{equation}\label{An}
 \sigma_K= \frac{A}{\omega_K N_0} \frac{4\pi}{\Delta\Omega\epsilon }
  \left[ S {\frac{d~}{{dE}} \left( \frac{N_{\rm x}}{N_{\rm e}} \right)} +
  \frac{N_{\rm x}}{N_{\rm e}} \mu_{\rm x} \frac{\cos\alpha}{\cos\beta} \right].
\end{equation}
When this method is used to determine the ionization cross section, error estimation becomes a difficult task. For example, Pérez \textit{et al.} \cite{perez15} gave estimations between 5 and 11\% for the $\sigma_K$ uncertainties in silicon, although no error was assumed in the two approximations abovementioned.

\subsection{Auger electrons}
The measurements of $\sigma_K$ based on Auger spectroscopy have been performed on gaseous targets. In this case, the intensity of Auger electrons can be written as 
$$N_A =\sigma_K N_{\rm e} \delta’ [\epsilon \Delta\Omega/4\pi] (1-\omega_K),$$
where $\epsilon \Delta\Omega/4\pi$ takes into account the detector efficiency, including the transmission of the analyzer, and $N_{\rm e}$, $\delta’$ and $\omega_K$ are defined above. Auger spectroscopy is particularly useful for determining $\sigma_K$ for atoms with inner-shell binding energies less than about 1 keV (with characteristic X-ray energies near or below the detection limits), and for which $\omega_K$ is small, \textit{i.e.}, the Auger yield ($1-\omega_K$)  is close to unity. 
The main sources of uncertainties in this method are the density $\delta’$ and the detection efficiency. Glupe and Melhorn \cite{glupe71} report total uncertainties of 5\% (3\% due to density, 2\% from counting statistics, and 2\% originated in the detection procedure). Similar uncertainties are reported by Hink \textit{et al.} \cite{hink81} for measurements with large enough overvoltage, but they increase when decreasing the overvoltage, down to $U=1.1$ or less. Platten \textit{et al.} \cite{platten85} and Quarles and Semaan \cite{quarles82} report uncertainties between 10 and 20\%.

\subsection{EELS}
Electron energy-loss spectroscopy is widely applied for thin-film analysis in transmission electron microscopy, and has been used for determining the ionization cross section of some elements. The energy spectrum of electrons transmitted reveals structures that can be associated with core-electron excitations,  similarly to X-ray absorption spectroscopy.
For a thin film of thickness $t$ and atomic density $\delta$, the relation between the measured intensity $N_E$ corresponding to energy losses associated with K-shell ionization
and the total transmitted intensity $N_T$ within a semiangle $\alpha$ is given by \cite{rossouw79}
$$N_{E}(\alpha)=N_{T}(\alpha) \delta t \sigma_K(\alpha)\; ,$$
where $\sigma_K(\alpha)$ is the partial ionization cross section, which tends to $\sigma_K$ when $\alpha$ approaches a critical value. 
There are  few works reporting measurements of $\sigma_K$ with this technique \cite{colliex72, isaacson72, egerton75, rossouw79}, with uncertainties estimated between 6 and 20\%. 

\section{\label{database}The K-shell ionization cross-section database}

The present database contains experimental K-shell ionization cross sections extracted from 103 publications covering the period 1930–2024, although no new experimental results were found after 2020. It includes 2509 data points, representing ionization cross-section values for 65 elements from H to U across the energy range $1.46\times 10^{-2}$~keV$<E<2\times 10^6$~keV. These data can be divided into three groups:
\begin{enumerate}
    \item The data reviewed in the paper published by Llovet \textit{et al.} \cite{llovet14}, which represents the most recent compilation. These data (amounting to 1953 data points for 63 elements) were retrieved from the original articlesyo.
    
    \item Results published prior to 2014 not included in  Llovet {\it et al.}'s work~\cite{llovet14}. This group comprises 382 values, most of which correspond to H and He ~\cite{smith30,liska34,harrison56,fite58, downey62, rothe62,asundi63,schram65,adamczyk66, schram66,gaudin67, shchemelinin75,brook78, nagy80}. Also, data for Ar, Ca, Ti, V, Mn, Ni, Cu, Y, Ag, Sn, Te, Pt and Au \cite{dolgov70, scholz72,dangerfield75,watanabe87, westbrook87, aydinol07} were found and integrated into the present database.  
    
    \item The measurements published after 2013, amounting 174 experimental values for Al,  Si, Ti, Fe, Ni, Zn, Zr, Ag, Te, Ta, Au and Bi \cite{fernandez14,barros15,perez15,mei16,vanin16,santos19,li20, tian20}.
\end{enumerate}
 
A summary of the information collected is shown in Table~\ref{publications}, ordered by atomic number $Z$. For each element, the energy range, the experimental method and target used, and the amount of data are recorded, along with the publication reference. To easily identify the publications, a key reference with the format AaXX is used, where Aa are the initial letters of the first author's surname, and the two numbers XX correspond to the publication year. The data compiled by Llovet {\it et al.}~\cite{llovet14} are grouped into a single entry for each element, and the key references corresponding to groups 2 and 3 are written in boldface.

The collected data are presented in Table 2\footnote{Not included in the arXiv version, see Supplementary data}, and also can be found in the supplementary material file. For each element, the table includes the following values: $E$, the corresponding overvoltage $U$ (defined as $U=E/I_K$, with $I_K$ the K-shell ionization energy~\cite{larkins77}),  $\sigma_K$, and the reported uncertainty $\Delta \sigma_K$. In the few cases for which $\Delta\sigma_K$ was not found, an estimate is given by considering the main sources of uncertainties involved in the corresponding experiment. Also, the method and target used for the determination, as well as the year of publication, are recorded. In addition, it is indicated whether the original work reports the ionization cross section or the X-ray production cross section. The fluorescence yield $\omega_K$ is included whenever it has been reported by the authors. When only $\sigma_K^X$ is provided, the $\sigma_K$ value was recovered by using the reported $\omega_K$.  For the cases where $\sigma_K^X$ is informed, but no $\omega_K$ value is given, the fluorescence yield was taken from Perkins {\it et al.} \cite{perkins91}. The last two columns of the table account for the reference key described above and to particular notes, clarifying certain entries.
 
A few discrepancies were found when comparing the recovered data with those presented by Llovet {\it et al.}~\cite{llovet14}. In the case of Westbrook and Quarles data \cite{westbrook87}, who measured $\sigma_K$ at 100 keV for a wide $Z$ range and present the results in graphical form without specifying the  targets used, the values retrieved by the present digitalization methodology for  V, Y, and Sn were assigned by Llovet {\it et al.}~\cite{llovet14} to  Ti, Zr and Sb,  respectively.  The data for  Ni, Ag, and Pt obtained from Scholz {\it et al.}~\cite{scholz72}, as well as those for Te from   Westbrook and Quarles~\cite{westbrook87} were not included in Llovet {\it et al.} compilation \cite{llovet14}. Similarly, data for Sn given by Motz and Placious~\cite{motz64} are missing in the tables and plots published by Llovet {\it et al.}~\cite{llovet14}, although the authors mention them in the main text. Also,  the work by Nagy {\it et al.} \cite{nagy80} for He is found neither in Table 2 nor in the  text  of Llovet {\it et al.}, although a reference key "Na80" in their corresponding plot appears to be linked to that author. 

The numerical data for the measurements of Clark~\cite{clark35}, Berkner~{\it et al.}~\cite{berkner70} and Scholz {\it et al.}~\cite{scholz72} were extracted from the tabulation given by Long {\it et al.}~\cite{long90}, who obtained the information by requesting it from the original authors. There are a few elements for which Scholz {\it et al.}~\cite{scholz72} reported two values obtained with different detectors; these data were treated in this work as independent determinations and their values were digitalized from the original paper to be included in the database, rather than using the single value provided by Long {\it et al.}~\cite{long90}

The dispersion and lack of experimental determinations for many elements and energy ranges hinder the systematization of the data through any form of analytical parameterization.

\section{\label{sec:discussion}Discussion}

\begin{figure}[h!]
    \centering
    \includegraphics[width=0.95\linewidth]{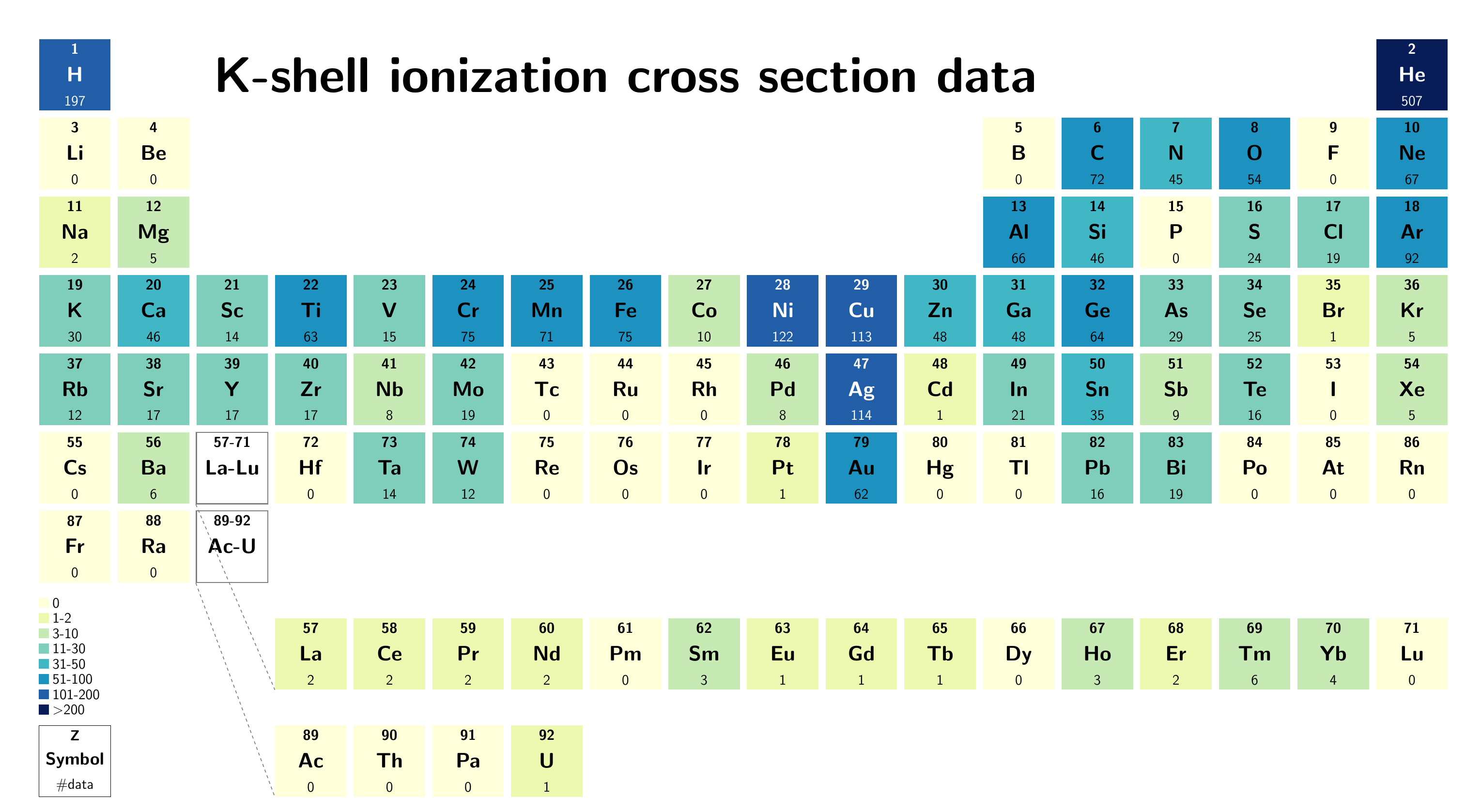}
    \caption{Compiled K-shell ionization cross section measurements for each element along the periodic table for $1\leq Z \leq 92$. The color scale identifies ranges of data numbers. The exact quantity of data found is given under the element's symbol.}
    \label{fig:periodictable}
\end{figure}

A panorama of the distribution of the compiled data across the periodic table is shown in Fig. \ref{fig:periodictable}, where the number of data points for each element is indicated. It can be seen that 27 elements have no data available in the considered range of atomic numbers, while 13 elements have only one or two data points, most of which reported by few authors \cite{scholz72,westbrook87,ishii77}. Conversely, around 52\% of the data correspond to just 8 elements: H (7.9\%), He (20.3\%), Ar (3.7\%), Cr (3\%), Fe (3\%), Ni (4.9\%), Cu (4.5\%), and Ag (4.6\%). The scarcity of experimental values for some third-period elements --particularly Na, Mg, and P--is noteworthy, despite the fact that X-ray determinations for these elements do not pose significant challenges in terms of X-ray detection, sample availability, or access to electron sources with adequate energies for K-shell ionization.

\begin{figure}
    \centering
\includegraphics[width=0.95\linewidth]{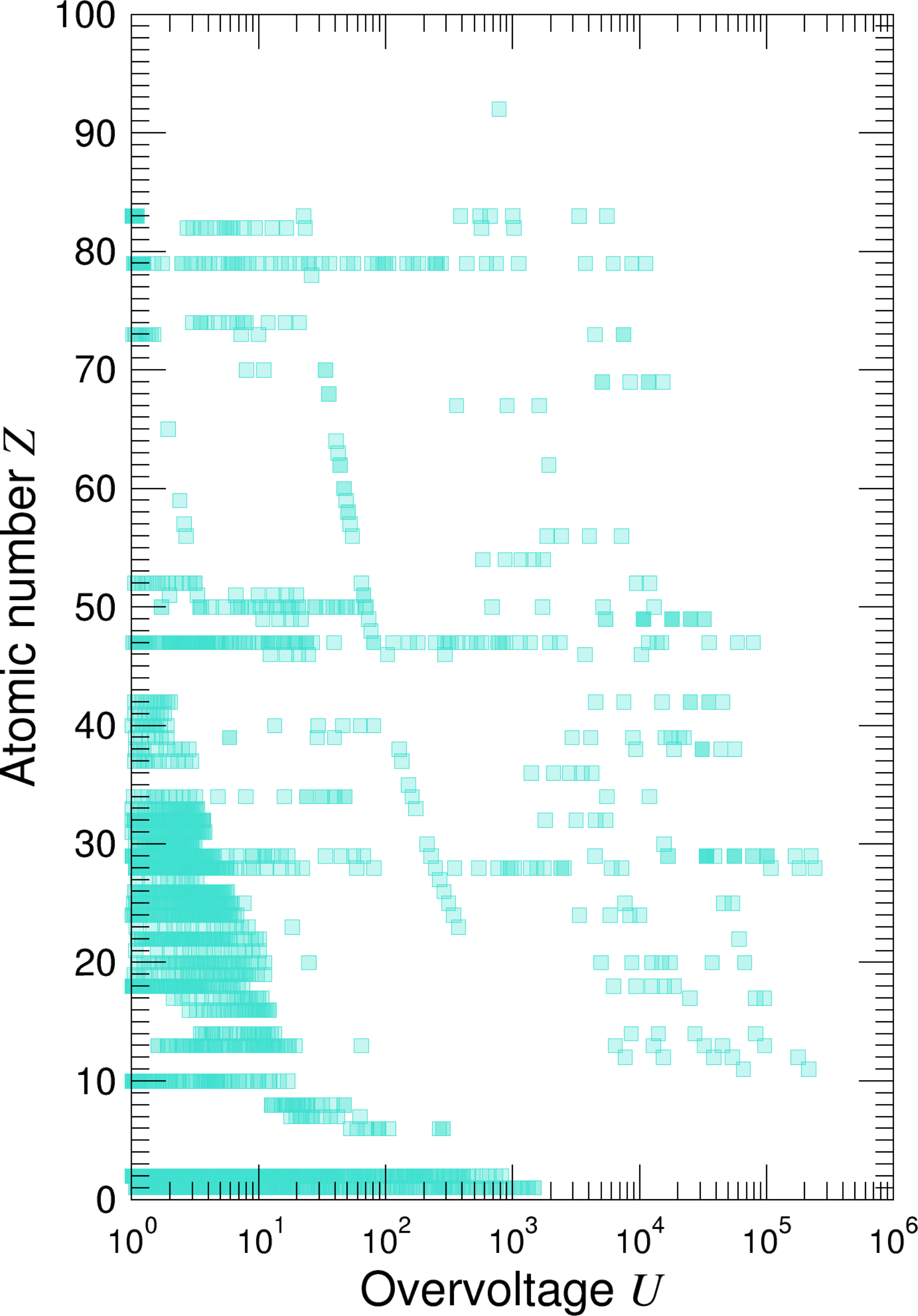}
    \caption{Distribution of the compiled measurements for K-shell ionization cross section for each atomic number as a function of the overvoltage $U$. Each square represents one data point.}
    \label{fig:ZvsU}
\end{figure}

Regarding the incident electron energies, it can be observed that over 50\% of the data are concentrated within a narrow energy range, spanning up to $U\approx 4$ (see Fig. \ref{fig:ZvsU}). Moreover, there are 695 data points with overvoltage $U<2$, \textit{i.e.}, almost 30\% of the total set. Notice that, for $E$ slightly above $I_K$, cross-section measurements have associated important uncertainties since, on the one hand, the X-ray emission is very weak due to the low ionization probability and, on the other hand, any error in the incident energy strongly affects $\sigma_K$ due to its rapid increase with energy.

Excluding the data for H and He, mainly obtained by counting ions or secondary electrons, the 91\% of the remaining measurements correspond to X-ray methods, which involve the use of the fluorescence yield $\omega_K$ to convert the number of detected photons into ionization cross section. This may introduce important uncertainties in the case of light elements (up to $Z\approx 10$) for which the $\omega_K$ values reported by different authors  present large discrepancies. For instance, for C  this parameter ranges from 0.0009 \cite{fink66} to 0.0035 \cite{hink71}. For heavier elements, the dispersion of the $\omega_K$ values is in general, lower than 10\%.

\begin{figure}[h!]
    \centering
    \includegraphics[width=0.95\linewidth]{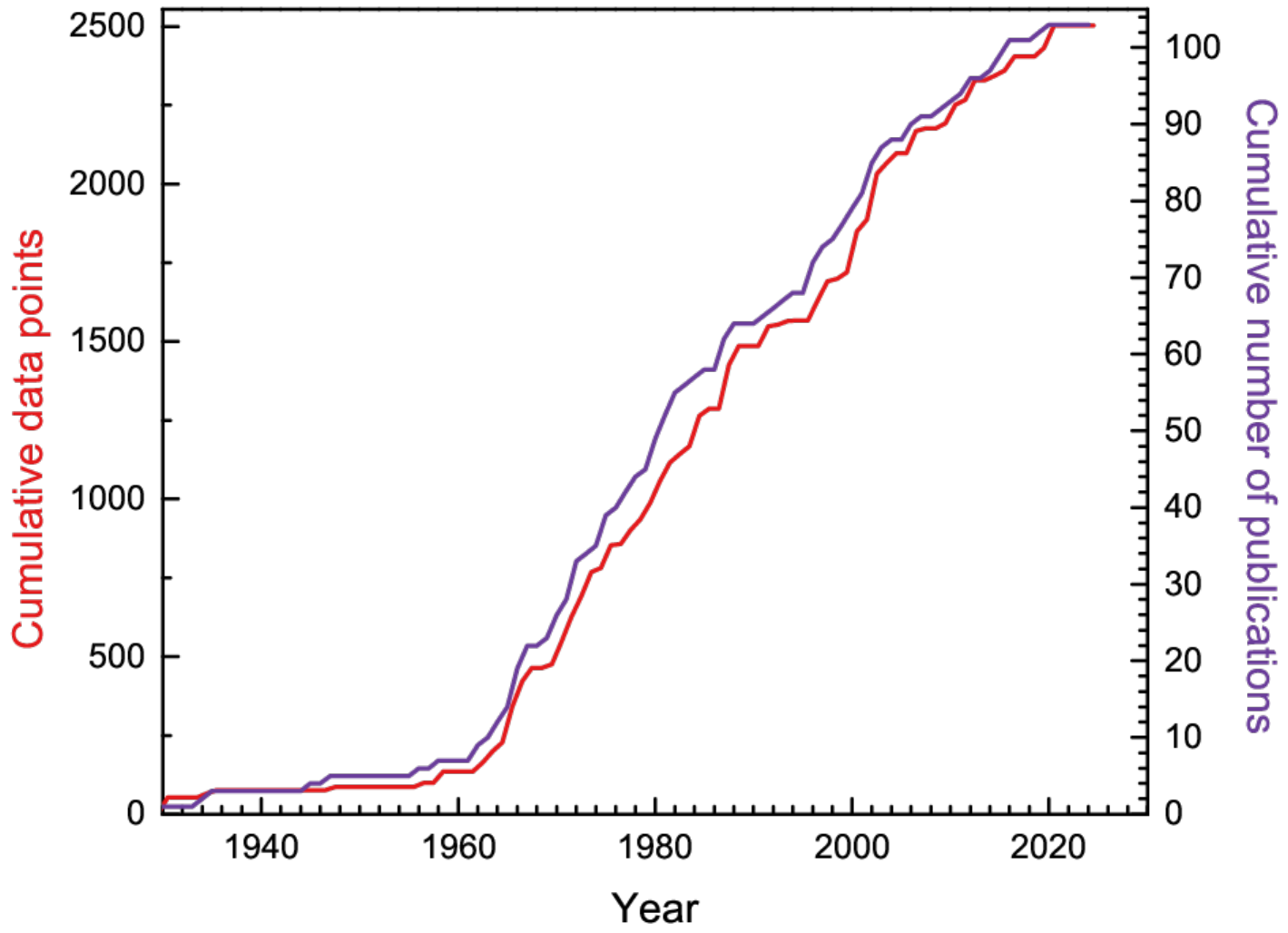}
     \caption{Cumulative number of $\sigma_K$ data points and publications along time.}
    \label{fig:timeevol}
\end{figure}

Analyzing the timeline of publications (see Fig. \ref{fig:timeevol}), the progression of experimental research on $\sigma_K$ reveals distinct phases. The initial stage, from 1930 to 1960, featured only a few experiments, primarily focused on H and He. This was followed by a more active period between 1965 and 1990, during which approximately 240 data points were measured, with an average of 10 articles published per lustrum. In the subsequent decade, the publication rate dropped by half, but it increased again during the 2000–2006 period. Since then, the pace of publications has slowed, although certain elements and energy ranges remain unexplored.

\begin{figure}
\includegraphics[width=0.5\columnwidth]{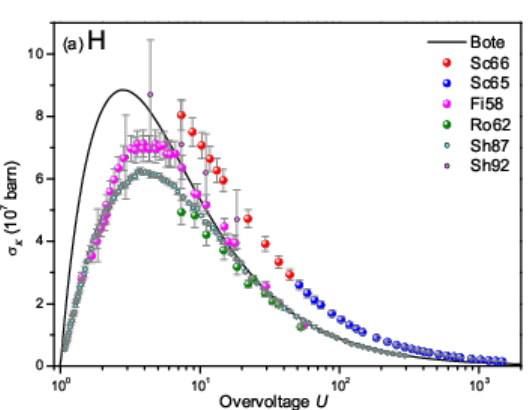}
\includegraphics[width=0.5\columnwidth]{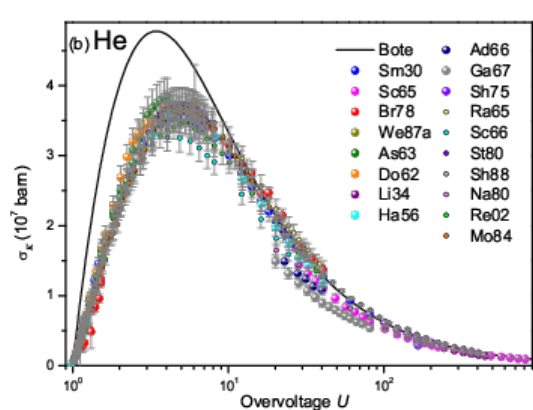}
\includegraphics[width=0.5\columnwidth]{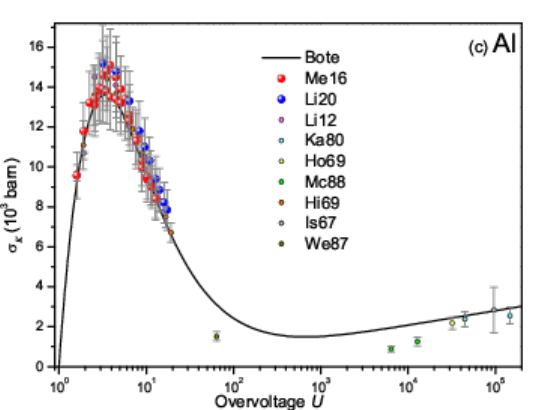}
\includegraphics[width=0.5\columnwidth]{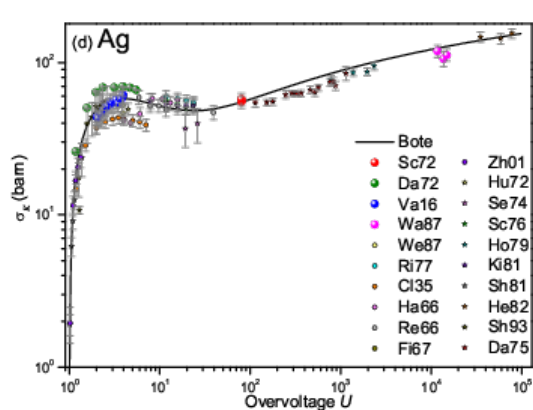}
\includegraphics[width=0.5\columnwidth]{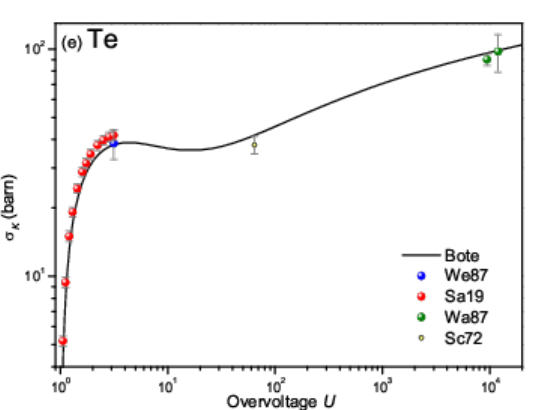}
\includegraphics[width=0.5\columnwidth]{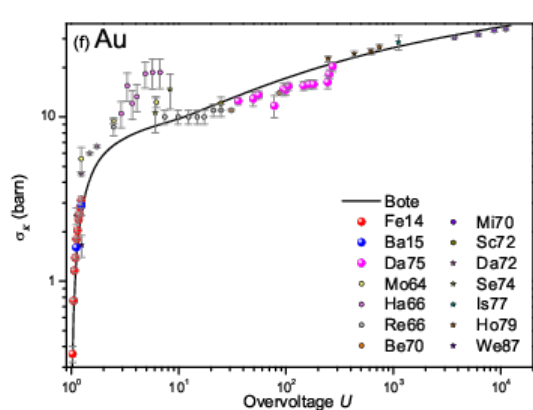}
\caption{\label{sigmas} Ionization cross section as a function of the overvoltage for different elements. The solid line represents the calculations performed by Bote \textit{et al}.\cite{bote09}. The experimental results not included in the compilation performed by Llovet \textit{et al.}\cite{llovet14} (groups 2 and 3 defined in \S\ref{database}) are shown as large spheres, and the remaining data (group 1) are plotted as small circles and stars. (a): hydrogen, (b): helium, (c): aluminum, (d): silver, (e): tellurium, and (f): gold.}
\end{figure}

Figure~\ref{sigmas} shows  the compiled data for $\sigma_K$ as a function of the overvoltage $U$ for selected elements. For comparison, predictions obtained with the parameterization given  by Bote \textit{et al.}~\cite{bote09} are included.  This  analytical model  results from fitting numerical data obtained with the distorted-wave Born approximation (DWBA), for $U\leq 16$, and the  plane wave Born approximation (PWBA) for higher overvoltages. 
It is worth noting that the values of $U$ associated with the experimental data were determined by using $I_K$ from a particular database \cite{larkins77}. According to the different tabulations for the ionization energy \cite{bearden67,lotz70,larkins77,nist},   differences in $I_K$ of the order  of $5$ eV are observed, which implies relative uncertainties not higher than 2\%. 

\onecolumn
\begin{table}[h!]
\caption{Measurements of K-shell ionization cross sections published up to 2024. Information on the incident electron energy range, method and target used, and reference is given. Methods include: measurements with X-ray yields (X), Auger yields (A), EELS spectra (E), ion number (I), and secondary electron number (SE). Targets used include self-supporting thin films (T), thin films on substrates (TS), thin films on thin substrates (TtS), thick substrates (S), and gases (G). {\it vm}: various methods; {\it vt}: various targets. The key for references AaXX is explained in the text and is compatible with the tables given in the supplementary material file. Data added in the present compilation (groups 2 and 3) are indicated with reference in bold typeface.
\label{publications}}
\end{table}
\setcounter{table}{0}
\vskip-0.5\intextsep
\setlength{\LTpre}{0pt}%
{
\sisetup{table-format=-2.5,  
    table-number-alignment=center, 
    table-space-text-pre =(, 
    table-space-text-post=\tc{$\star$$\star$$\star$}, 
    table-align-text-pre=false, 
    table-align-text-post=false}
\begin{longtable}{ccccc}
\hline \hline
\multicolumn{1}{c}{\textbf{Element}} & \multicolumn{1}{c}{\textbf{Energy range [keV]}} & \multicolumn{1}{c}{\textbf{Method/Target}}  & \multicolumn{1}{c}{\textbf{Number of data}}  & \multicolumn{1}{c}{\textbf{Refs.}} \\ 
\hline 
\endfirsthead
\multicolumn{5}{c}%
{{\bfseries \tablename\ \thetable{} -- continued from previous page}} \\
\hline 
\multicolumn{1}{c}{\textbf{Element}} & \multicolumn{1}{c}{\textbf{Energy range [keV]}} & \multicolumn{1}{c}{\textbf{Meth./Tar.}}  & \multicolumn{1}{c}{\textbf{Data points}}  & \multicolumn{1}{c}{\textbf{Refs.}}  \\ 
\hline 
\endhead

\hline  \multicolumn{5}{r}{{Continued on next page}} \\ 
\endfoot
\endlastfoot
\hline
H & 0.014--4  & {\it vm}/G & 113 & Sh87~\cite{shah87}, Sh92\cite{shyn92}\\
H & 0.019--0.752 & I/G & 37 & \textbf{\textbf{Fi58}}~\cite{fite58} \\
H & 0.100--0.718 & I/G & 11& \textbf{\textbf{\textbf{Ro62}}}~\cite{rothe62} \\
H & 0.7-20   & I/G & 26 & \textbf{\textbf{\textbf{Sc65}}}~\cite{schram65} \\
H & 0.1--0.6  & I/G & 10 &  \textbf{\textbf{Sc66}}~\cite{schram66} \\
He & 0.02--1.6 & I/G & 291 & Ra65~\cite{rapp65}, {Sc66}~\cite{schram66}, Na80~\cite{nagy80}, St80~\cite{stephan80},  \\
   &   &   &   &   Mo84~\cite{montague84},We87a~\cite{wetzel87}, Sh88~\cite{shah88}, Re02~\cite{rejoub02} \\
He & 0.025--4.5 & I/G & 54 & \textbf{{Sm30}}~\cite{smith30} \\
He & 0.7--20 & I/G & 26 &  \textbf{{Sc65}}~\cite{schram65}\\
He & 0.027--0.997 & I/G & 27 & \textbf{{Br78}}~\cite{brook78}\\
He & 0.025--0.1 & I/G & 36 & \textbf{As63} \cite{asundi63}\\
He & 0.025--0.08 & I/G & 18 &  \textbf{Do62} \cite{downey62}\\
He & 2.5--11 & I/G & 12 &  \textbf{Li34} \cite{liska34}\\
He & 0.024--1 & I/G & 12 &  \textbf{Ha56} \cite{harrison56}\\
He & 0.566--0.984 & I/G & 5 & \textbf{Ad66} \cite{adamczyk66}\\
He & 0.5--2 & I/G & 14 & \textbf{Ga67} \cite{gaudin67}\\
He & 4.085 & I/G & 1 & \textbf{Sh75} \cite{shchemelinin75}\\
C  & 0.29--80 & {\it vm}/{\it vt} & 72 & Gl67 \cite{glupe67}, Hi71~\cite{hink71}, Co72~\cite{colliex72}, Is72~\cite{isaacson72}, \\
   &   &   &   &Ta73~\cite{tawara73}, Eg75~\cite{egerton75}, Ro79~\cite{rossouw79}, Li12~\cite{limandri12}\\
N  & 0.85--25 &  {\it vm}/{\it vt} & 45 & Gl71~\cite{glupe71}, Is72~\cite{isaacson72}, Ta73~\cite{tawara73} \\
O  & 1--25    & {\it vm}/{\it vt} & 54 & Gl71~\cite{glupe71}, Is72~\cite{isaacson72}, Ta73~\cite{tawara73}, Pl85~\cite{platten85}, \\
   &     &   &  &Li12~\cite{limandri12} \\
Ne & 0.9--14.6 & {\it vm}/G& 67 & Gl71~\cite{glupe71}, Ta73~\cite{tawara73}, Pl85\cite{platten85}, Hi81~\cite{hink81}\\
Na & 7$\times 10^4$--$2.3\times 10^5$ & X/T & 2 & Ka80~\cite{kamiya80} \\
Mg & $1\times 10^4$--$2.3\times 10^5$ & X/T & 5 & Ho79~\cite{hoffmann79}, Ka80~\cite{kamiya80}, Mc88~\cite{mcdonald88} \\
Al & 2.5--2.3$\times 10^5$& {\it vm}/T & 26 & Hi69~\cite{hink69}, Is77~\cite{ishii77}, Ho79~\cite{hoffmann79}
, Ka80~\cite{kamiya80},\\ 
   &   &   &   & {We87b}~\cite{westbrook87}, Mc88\cite{mcdonald88}, Li12~\cite{limandri12}\\
Al & 4--20 & X/TS & 30 & \textbf{Me16}~\cite{mei16} \\
Al & 5--27 & X/S & 10 & \textbf{Li20}~\cite{li20} \\
Si & 2.5--1.5$\times 10^5$ & {\it {\it vm}}/{\it vt} & 32 & Is77~\cite{ishii77}, Ho79~\cite{hoffmann79}, Pl85~\cite{platten85}, Sh94~\cite{shchagin94}, \\
   &   &    &    & Zh09~\cite{zhu09}, Li12~\cite{limandri12}  \\
Si & 2.1--20 &  X/S & 14 &\textbf{\textbf{Pe15}} \cite{perez15}\\
S  & 7--30 &  X/TS  & 24 & Wu10~\cite{wu10} \\
Cl & 6--2.7$\times 10^5$  & X/{\it vt} & 19  & Is77~\cite{ishii77}, Ka80~\cite{kamiya80}, Wu11~\cite{wu11} \\
Ar & 3.2--6$\times 10^4$ & X/G & 75 & Ta73~\cite{tawara73}, Ho79~\cite{hoffmann79}, Qu82~\cite{quarles82}, Hi82~\cite{hippler82} \\
    &    &    &    & Hi83\cite{hippler83}, Pl85\cite{platten85}, Si03\cite{singh03} \\
Ar & 3.5--11 & X/G  & 9 &  \textbf{ \textbf{Ay07}}~\cite{aydinol07} \\
Ar & 4--11 & X/G & 8 & \textbf{\textbf{Do70}}~\cite{dolgov70}\\
K  & 3.75--45 &  X/{\it vt} & 30 & Sh91~\cite{shevelko91}, Wu12~\cite{wu12} \\
Ca & 4.5--2.7$\times 10^5$ &  X/{\it vt} & 45 & Is77~\cite{ishii77}, Ho79~\cite{hoffmann79}, Sh91~\cite{shevelko91}, Wu10~\cite{wu10} \\
Ca & 100 & X/T & 1 & \textbf{We87b}~\cite{westbrook87} \\
Sc & 4.8--45 & X/TS & 14 & An00\cite{an00}  \\
Ti & 5.6--50 &  X/{\it vt} & 53 & Je75\cite{jessenberger75}, He97\cite{he97}, An03\cite{an03}, Li12\cite{limandri12}  \\
Ti & $3\times 10^5$ & X/T & 1 & \textbf{Wa87}\cite{watanabe87} \\
Ti & 7--27  & X/S & 9 & \textbf{Li20} \cite{li20} \\
V  & 5.9--2$\times 10^3$ &  X/{\it vt} & 14 & Sc72\cite{scholz72}, An00\cite{an00} \\
V  & 100 & X/T & 1 & \textbf{{We87b}}~\cite{westbrook87} \\
Cr & 6--6$\times 10^4$ &  X/{\it vt} & 75 & Sc72~\cite{scholz72}, Ho79~\cite{hoffmann79}, Lu96~\cite{luo96}, He97~\cite{he97},\\
   &   &   &   & Ll00~\cite{llovet00}, An03~\cite{an03} \\
Mn & 3.5--5$\times 10^4$ &  X/{\it vt} & 69 & Fi67~\cite{fisher67}, Sc72~\cite{scholz72}, Ho79~\cite{hoffmann79}, Sh80~\cite{shima80},\\
   &   &   &   &Lu97~\cite{luo97}, Ta99b~\cite{tang99b}, Ll02~\cite{llovet02} \\
Mn & $3\times 10^5$--$3.5 \times 10^5$ & X/T & 2 & \textbf{Wa87}~\cite{watanabe87} \\
Fe & 7.5--2$\times 10^3$ &  X/{\it vt}   & 59  & Sc72~\cite{scholz72}, He96a~\cite{he96a}, Lu97~\cite{luo97}, Ll02~\cite{llovet02} \\
Fe & 8.3--28 & X/T & 16 & \textbf{Ti20}~\cite{tian20} \\
Co & 8.5--2$\times 10^3$ &  X/{\it vt} & 10 & Sc72~\cite{scholz72}, An96~\cite{an96}\\
Ni & 8.9--2$\times 10^6$ &  X/{\it vt} & 94 & Sm45~\cite{smick45}, Po47~\cite{pockman47}, Se74~\cite{seif74}, Je75~\cite{jessenberger75}, \\
   &   &   &   & Ho79~\cite{hoffmann79}, Ge82~\cite{genz82}, Lu96~\cite{luo96}, He97~\cite{he97},\\
   &   &   &   &Ll00~\cite{llovet00}, An06~\cite{an06} \\
Ni & 2$\times 10^3$ &  X/T & 1 & \textbf{Sc72} \cite{scholz72} \\
Ni &  $2.95\times 10^3$--$2.9 \times10^4$ &  X/T & 11 & \textbf{Da75} \cite{dangerfield75} \\
Ni & 9.3--28.2 & X/T & 16 & \textbf{Ti20} \cite{tian20} \\
Cu & 9--2$\times 10^6$ & X/{\it vt} & 111 & Fi67~\cite{fisher67}, Mi70~\cite{middleman70}, Da72~\cite{davis72}, Hu72~\cite{hubner72},\\
   &    &   &   & Sc72~\cite{scholz72}, Is77~\cite{ishii77}, Be78~\cite{berenyi78}, Ho79~\cite{hoffmann79}, \\ 
   &   &   &   &  Sh80~\cite{shima80}, Sh81~\cite{shima81}, Ge82~\cite{genz82}, {{We87b}}~\cite{westbrook87},\\ 
   &   &   &   & An96~\cite{an96}, He97~\cite{he97}, Ll00~\cite{llovet00}, Zh01~\cite{zhou01}\\
Cu & $3\times 10^5$--$3.5 \times 10^5$ & X/T & 2 & \textbf{Wa87}~\cite{watanabe87} \\
Zn & 10.3--1.5$\times 10^5$ & X/{\it vt} & 32 & Sc72~\cite{scholz72}, Is77~\cite{ishii77}, Ta99a~\cite{tang99a}, Wu10~\cite{wu10}  \\
Zn & 11.1--28.6 & X/T & 16 & \textbf{Ti20} \cite{tian20}\\
Ga & 10.5--39 & X/{\it vt} & 48 & Zh01~\cite{zhou01a}, Zh02~\cite{zhou02}, Me06~\cite{merlet06} \\
Ge & 11.2--6$\times 10^4$  & X/{\it vt}& 64 & Ho79~\cite{hoffmann79}, Sh81~\cite{shima81}, Ta02~\cite{tang02}, Zh02~\cite{zhou02}, \\
   &   &   &   & Me04~\cite{merlet04}, Lu01~\cite{luo01} \\
As & 12--2$\times 10^3$ & X/T & 29 & Sc72~\cite{scholz72}, Me06~\cite{merlet06} \\
Se & 13--1.5$\times 10^5$ & X/{\it vt}& 25 & Fi67~\cite{fisher67}, Sc72~\cite{scholz72}, Is77~\cite{ishii77}, Be78~\cite{berenyi78},\\ 
   &   &   &   & Ki81~\cite{kiss81}, Lu01~\cite{luo01}]  \\
Br & 2$\times 10^3$ & X/T & 1 &  [Sc72~\cite{scholz72} \\
Kr & 2$\times 10^4$--6$\times 10^4$ & X/G & 5 & Ho79~\cite{hoffmann79} \\
Rb & 16--2$\times 10^3$ & X/{\it vt} & 12 & Sc72~\cite{scholz72}, Sh91~\cite{shevelko91} \\
Sr & 17--9$\times 10^3$ & X/{\it vt} & 17 & Mi70~\cite{middleman70}, Sc72~\cite{scholz72}, Sh91~\cite{shevelko91} \\
Y & 17--2.7$\times 10^5$
& X/{\it vt} & 14 & Se74~\cite{seif74}, Is77~\cite{ishii77}, Ho79~\cite{hoffmann79}, Lu01~\cite{luo01} \\
Y & $3\times 10^5$--$3.8 \times 10^5$ & X/TS & 3 & \textbf{Wa87}~\cite{watanabe87} \\
Y &  $100$       &   X/T &        2 &           \textbf{We87b}~\cite{westbrook87}\\
Zr & 18--1.4$\times 10^3$ &X/{\it vt} & 13 & Ha64~\cite{hansen64}, Zh02~\cite{zhou02}\\
Zr & 20--27 & X/TS  & 4 & \textbf{Li20}~\cite{li20} \\
Nb & 20--34 & X/TS  & 8 & Pe98~\cite{peng98} \\
Mo & 21--9$\times 10^5$ & X/{\it vt} & 19 & Mi70~\cite{middleman70}, He96b~\cite{he96b}, Ri77~\cite{ricz77} \\
Pd & 300--7.1$\times 10^5$ & X/T & 8 & Be70~\cite{berkner70}, Is77~\cite{ishii77}, Ri77~\cite{ricz77} \\
Ag & 26--2$\times 10^6$ 
& X/{\it vt} & 80 & Cl35~\cite{clark35}, Ha66~\cite{hansen66}, Re66~\cite{rester66}, Fi67~\cite{fisher67}, \\
    &   &   &   & Da72~\cite{davis72}, Hu72~\cite{hubner72}, Se74~\cite{seif74}, Sc76~\cite{schlenk76}, \\
    &   &   &   & Ri77~\cite{ricz77}, Ho79~\cite{hoffmann79}, Ki81~\cite{kiss81}, Sh81~\cite{shima81}, \\
    &   &   &   & Ge82~\cite{genz82}, {We87b}~\cite{westbrook87}, Sc93~\cite{schneider93}, Zh01~\cite{zhou01} \\
Ag & 2040 & X/T & 2 & \textbf{Sc72} \cite{scholz72} \\
Ag &  $2.9\times 10^3$--$2.9 \times10^4$ &  X/T & 15  & \textbf{Da75}~\cite{dangerfield75} \\
Ag & $3\times 10^5$--$3.8 \times 10^5$ & X/T & 3 & \textbf{Wa87}~\cite{watanabe87} \\
Ag & 51.5--100.8 & X/T & 14 & \textbf{Va16} \cite{vanin16}  \\
Cd & 2040 & X/T & 1 & Sc72~\cite{scholz72} \\
In & 300--9$\times 10^5$ & X/T & 21 & Mi70~\cite{middleman70}, Sc72~\cite{scholz72}, Is77~\cite{ishii77}, Ri77~\cite{ricz77} \\
Sn & 50--1.5$\times 10^5$ & X/T & 29 & Fi67~\cite{fisher67}, Ha64~\cite{hansen64}, Re66~\cite{rester66}, Ha66~\cite{hansen66}, \\
   &   &   &   & Sc72~\cite{scholz72}, Ri77~\cite{ricz77}, Is77~\cite{ishii77}, Ho79~\cite{hoffmann79} \\
Sn & 50--500 & X/T & 4 & \textbf{Mo64} \cite{motz64} \\
Sn & $3.8 \times 10^5$ & X/TS & 1 & \textbf{Wa87}~\cite{watanabe87} \\
Sn & 100 & X/T & 1 & \textbf{We87b}~\cite{westbrook87} \\
Sb & 60--2$\times 10^3$ & X/T & 9 & Sc72~\cite{scholz72}, Ki81~\cite{kiss81} \\
Te & 2$\times 10^3$ & X/T & 1 & Sc72\cite{scholz72}  \\
Te & $3 \times 10^5$--$3.8 \times 10^5$ & X/TS & 2 & \textbf{Wa87}~\cite{watanabe87} \\
Te & 100 & X/T & 1 & \textbf{We87b}~\cite{westbrook87} \\
Te & 33--100 & X/T & 12 & \textbf{Sa19} \cite{santos19} \\
Xe & 2$\times 10^4$--6$\times 10^4$ & X/G & 5 & Sc72~\cite{scholz72} \\
Ba & 100--2.7$\times 10^5$ & X/T & 6 & Sc72~\cite{scholz72}, Is77~\cite{ishii77}, {We87b}~\cite{westbrook87}] \\
La & 100; 2$\times 10^3$ & X/T & 2 & [Sc72~\cite{scholz72}, We87b~\cite{westbrook87} \\
Ce & 2$\times 10^3$ & X/T & 2 & Sc72~\cite{scholz72} \\
Pr & 100; 2$\times 10^3$  & X/T & 2 & Sc72~\cite{scholz72}, {We87b}~\cite{westbrook87} \\
Nd & 2$\times 10^3$ & X/T & 2 & Sc72~\cite{scholz72}\\
Sm & 2$\times 10^3$; 9$\times 10^4$  & X/T & 3 & Sc72~\cite{scholz72}, Is77~\cite{ishii77} \\
Eu & 2$\times 10^3$ & X/T & 1 & Sc72~\cite{scholz72}\\
Gd & 2$\times 10^3$ & X/T & 1 & Sc72~\cite{scholz72} \\
Tb & 100 & X/T & 1 & {We87b}~\cite{westbrook87} \\
Ho & 2$\times 10^4$--9$\times 10^4$ & X/T & 3 & Sc72~\cite{scholz72}, Is77~\cite{ishii77} \\
Er & 2$\times 10^3$ & X/T & 2 & Sc72~\cite{scholz72} \\
Tm & 3$\times 10^5$--9$\times 10^5$ & X/T & 6 & Mi70~\cite{middleman70}  \\
Yb & 490--2$\times 10^3$ & X/T & 4 & Sc72~\cite{scholz72}, Se74~\cite{seif74} \\
Ta & 490--5$\times 10^5$ & X/T & 5 & Mi70~\cite{middleman70}, Se74~\cite{seif74} \\
Ta & 68--100 & X/T & 9 & \textbf{Sa19} \cite{santos19} \\
W  & 210--1.4$\times 10^3$ & X/T & 12 & Ha64~\cite{hansen64},Ha66~\cite{hansen66} \\
Pt & 2040 & X/T & 1 & \textbf{Sc72}~\cite{scholz72} \\
Au & 90--9$\times 10^6$ & X/T & 37 & Mo64~\cite{motz64}, Ha66~\cite{hansen66}, Re66~\cite{rester66}, Be70~\cite{berkner70}, \\
   &   &   &   & Mi70~\cite{middleman70}, Sc72~\cite{scholz72}, Da72~\cite{davis72}, Se74~\cite{seif74}, \\
   &   &   &   & Is77~\cite{ishii77}, Ho79~\cite{hoffmann79},{We87b}~\cite{westbrook87}  \\
Au &  $2.9\times 10^3$--$2.9 \times10^4$ &  X/T & 13  & \textbf{Da75} \cite{dangerfield75} \\
Au & 82--100 & X/T & 10 & \textbf{Fe14} \cite{fernandez14} \\
Au & 89--100 & X/T & 2 & \textbf{Ba15} \cite{barros15} \\
Pb & 240--9$\times 10^4$ & X/T & 16 & Ha64~\cite{hansen64}, Ha66~\cite{hansen66}, Sc72~\cite{scholz72}, Se74~\cite{seif74}, \\ 
   &   &   &   & Is77~\cite{ishii77}, Ho79~\cite{hoffmann79}  \\
Bi & 2$\times 10^3$--5$\times 10^5$ & X/T & 7 & Mi70~\cite{middleman70}, Sc72~\cite{scholz72}, Is77~\cite{ishii77}, Ho79~\cite{hoffmann79}\\
Bi & 92--100 & X/T & 5 & \textbf{Fe14} \cite{fernandez14}\\
Bi & 91--100 & X/T & 7 & \textbf{Sa19} \cite{santos19} \\
U  & 9$\times 10^4$ & X/T & 1 & Is77~\cite{ishii77} \\
\hline \hline
\end{longtable}
}
    

\twocolumn

For certain elements, a great dispersion is observed between different authors. However, in some cases the experiments present the same global behavior, similar to that described by the theoretical predictions. For instance, the energy of the local maximum for H cross section (at  $U\approx 4$) can be clearly deduced from the data, although the $\sigma_K$ values vary within 35\% (see Fig. \ref{sigmas}a). Another illustrative case is Ni, with a dispersion of 37\% at $U\approx 3.6$. In other cases, determinations from different authors do not follow a clear trend. This is the case of Au for certain energy ranges (Fig. \ref{sigmas}f), and other elements such as Sn and Pb. For some elements, instead, the experimental results present low dispersion; such is the case of Al illustrated in Fig. \ref{sigmas}c). There are elements for which the compilation performed by Llovet \textit{et al.} \cite{llovet14} contains very few data, but with the values included in the present work it is possible to observe a definite behavior; such is the case of Te (Fig. \ref{sigmas}e).

Regarding the relationship between the experimental and theoretical data, different situations can be pointed out. For H (Fig. \ref{sigmas}a), there is a large amount of experimental data that show important discrepancies between the different sources. In this case, the theoretical assessment overestimates all of them for overvoltages below 3, \textit{i.e.}, in the curve region at the left side of the maximum. For higher energies, the data published by Schram \textit{et al.}\cite{schram65,schram66} are quite above the general trend of both theoretical and other experimental data. For He (Fig. \ref{sigmas}b), many data were found, which present much smaller discrepancies than for H. As can be seen, the theoretical predictions are in good agreement with experiments for $U$>10, but seriously overestimate the data cloud for lower overvoltages. For S, only one set of measurements was found, which clearly differs from the DWBA formalism. With regard to Ag (Fig. \ref{sigmas}d), Au (Fig. \ref{sigmas}f), and Ni, the different experimental data sets present large discrepancies between them. However, the theoretical results follow the general trend. In the particular case of Au, some of the measurements show a strange behavior.

\section{\label{sec:conclusions}Conclusions}

The experimental K-shell ionization cross-sections by electron impact available in the literature up to 2024 were compiled to build a new database, increasing the number of previously compiled data in around 29\%.

In addition, the different experimental methods used to determine the ionization cross-sections were discussed in detail, along with the main sources of uncertainties.

The compiled data present an important dispersion for several elements, in some cases even greater than the corresponding uncertainties, which are also included in the present dataset.

This compilation of K-shell ionization cross-section data allowed to perform a statistical analysis, which showed the lack of data for certain elements or energy regions. Therefore, this analysis can serve as a guide for researchers to perform new experimental determinations.

\section{\label{sec:suppldata}Supplementary data}
The database is also openly available in 

\href{https://www.famaf.unc.edu.ar/~trincavelli/database.html}{https://www.famaf.unc.edu.ar/~trincavelli/database.html}.

\onecolumn

\twocolumn

\section*{\label{sec:declaration}Declaration of competing interest}
The authors declare that they have no known competing financial interests or personal relationships that could have appeared to
influence the work reported in this paper.


\section*{\label{sec:ackn}Acknowledgments}

The authors acknowledge the following institutions of Argentina for financial support:
the Consejo Nacional de Investigaciones científicas y Técnicas (CONICET), by the projects PIP 11220200100986CO and PIP 11220200102421CO, the Agencia Nacional de Promoción Científica y Tecnológica (ANPCyT), by the project PICT-2020-SERIE A-01931, and the Secretaría de Ciencia y Técnica de la Universidad Nacional de Córdoba, by the project 33620230100222CB.

\nocite{*}
\bibliographystyle{elsarticle-num} 
\bibliography{references}
\end{document}